# Stochastic parallel gradient descent based adaptive optics used for high contrast imaging coronagraph

Bing Dong[1], Deqing Ren[1,2], Xi Zhang[2]

[1]Physics & Astronomy Department, California State University Northridge, California, 91330-8268, USA; bdong@csun.edu

[2]National Astronomical Observatories/Nanjing Institute of Astronomical Optics and Technology, Chinese Academy of Sciences, Nanjing 210042, China

**Abstract**

An adaptive optics (AO) system based on stochastic parallel gradient descent (SPGD) algorithm is proposed to reduce the speckle noises in the optical system of stellar coronagraph in order to further improve the contrast. The principle of SPGD algorithm is described briefly and a metric suitable for point source imaging optimization is given. The feasibility and good performance of SPGD algorithm is demonstrated by experimental system featured with a 140-actuators deformable mirror (DM) and a Hartmann- Shark wavefront sensor. Then the SPGD based AO is applied to a liquid crystal array (LCA) based coronagraph. The LCA can modulate the incoming light to generate a pupil apodization mask in any pattern. A circular stepped pattern is used in our preliminary experiment and the image contrast shows improvement from $10^{-3}$ to $10^{-4.5}$ at angular distance of $2\lambda/D$ after corrected by SPGD based AO.

**Key Words**：instrumentation: adaptive optics, methods: laboratory, techniques: image processing

## 1 INTRODUCTION

Exoplanet direct imaging and characterization has attracted much more attention in recent years, which may help to answer one of the most fundamental scientific questions like "are we alone in the universe". To directly image an Earth-like planet orbiting a nearby star in the visible, we have to employ an extremely high-contrast stellar coronagraph at a level of $10^{-10}$ at small angular separations with a space telescope (a contrast of $10^{-6}$ to $10^{-7}$ is needed in the infrared). Many approaches have been proposed to achieve $10^{-10}$ contrast theoretically within inner working angle of $5\lambda/D$ (Guyon et al. 2006a). However, to obtain this high contrast is still very challenging at the state of the art as there are many factors impacting its real performance. The most essential limitation of coronagraph is the speckle noises near the star arising from the imperfections of the optical system and may completely overwhelm any faint planet images. So extremely-high quality of wavefront is required in high contrast coronagraph.

AO has been considered as an effective way to eliminate the impact of background speckles in coronagraph (Malbet et al. 1995, Give'on et al. 2007). In recent years, the AO system employing a high actuator-density DM and focal-plane wavefront sensing like phase-retrieval is demonstrated successfully both in laboratory and on ground-based telescopes to eliminate speckle noises in coronagraph (Trauger & Traub 2007, Serabyn et al. 2010). Focal-plane wavefront sensing is immune to non-common path errors and chromaticity and also simplifies the optical system, so it is optimal for high contrast imaging (Guyon et al. 2006b). However, the existing approaches are highly dependent on high accuracy estimation of the pupil wavefront (about $\lambda/2000$ rms) and high accuracy positioning of the DM actuators (less than 0.2nm) and related wavefront sensing algorithms are somewhat complicated. In this paper, we consider using a focal-plane wavefront sensing method named as SPGD to reduce the speckle noises in coronagraph. Without an estimation of the wavefront, the DM is driven directly by measurements in the image plane in SPGD. SPGD is also a model-free method so it is very robust to model errors of DM. After first developed by Vorontsov (Vorontsov & Carhart 1997), SPGD method has been successfully used in many scenarios (Weyrauch & Vorontsov 2005, Zhou et al. 2009). As far as we know, it has not been used for high contrast imaging coronagraph.

We are particularly interested in one kind of pupil-apodized coronagraph using a step-transmission filter (Ren & Zhu 2007, Ren et al. 2010). Recently, Ren & Zhu firstly proposed a high-contrast imaging coronagraph that integrates a liquid crystal array (LCA) and a DM for active pupil apodization and phase correction respectively (Ren & Zhu 2011). Avoiding the difficulty of fabricating a precise filter mask, this new approach is low-cost and flexible to work with any type of telescope.

Nevertheless, the wavefront errors in the optical system especially the surface error of the LCA is still our concern and must be suppressed by AO. In our case, AO is mainly used to correct quasi-static errors of the optical system so a high close-loop bandwidth is not necessary.

In Section 2, we briefly present the principle of SPGD algorithm and give a metric suitable for point source imaging optimization. Then an experimental system is set up to demonstrate the feasibility and effectiveness of SPGD algorithm in Section 3. The SPGD based AO is applied to a LCA based coronagraph to improve the image contrast and the result is given. Finally, we present our conclusions.

## 2 PRINCIPLE OF SPGD ALGORITHM

SPGD algorithm is an improved version of the well-known steepest descent algorithm. It applies small random perturbations to all control parameters (voltages of actuators) simultaneously, and then evaluates the gradient variation of system performance metric (J). The wavefront errors can be compensated by optimizing the metric that is calculated from measurable data, like intensity distributions, of the focal plane. The control signals update in iteration process of SPGD algorithm as following rule:

$$\mathbf{u}^{k+1} = \mathbf{u}^k - \gamma \delta J^k \delta \mathbf{u}^k \qquad (1)$$

where $k$ is the iteration number; $\mathbf{u} = \{u_1, u_2, \ldots, u_N\}$ is the control signal vector, N is the control channel number (i.e. the number of actuators); $\gamma$ is the gain coefficient which is positive for minimizing the metric and negative for maximizing the metric; $\delta \mathbf{u}$ denotes small random perturbations that have identical amplitudes and Bernoulli probability distribution; $\delta J$ is the variation of system performance metric.

$$\delta J = J(\mathbf{u} + \delta \mathbf{u}) - J(\mathbf{u}) = J(u_1 + \delta u_1, \ldots, u_j + \delta u_j, \ldots u_N + \delta u_N) - J(u_1, \ldots, u_j, \ldots u_N) \qquad (2)$$

To improve the estimation accuracy of $\delta J$, two-sided perturbation is used as:

$$\delta J = J_+ - J_- = J(\mathbf{u} + \delta \mathbf{u}/2) - J(\mathbf{u} - \delta \mathbf{u}/2) \qquad (3)$$

The gain coefficient $\gamma$ adaptive to the metric J is used to accelerate the convergence.

$$\gamma^{k+1} = \gamma^k \cdot J^k \quad \text{(to minimize J)} \qquad (4)$$

or

$$\gamma^{k+1} = \gamma^k / J^k \quad \text{(to maximize J)} \qquad (5)$$

In our case, a point source will be imaged on the focal plane and several metrics can be used to evaluate the system performance (Chen et al. 2009). Here the mean radius of image spot is chosen as the performance metric to be minimized:

$$J = \frac{\iint \sqrt{(x-x')^2 + (y-y')^2} I(x,y) dxdy}{\iint I(x,y) dxdy} \quad (6)$$

where $I(x,y)$ is the intensity distribution in the focal plane; $(x', y')$ is the image centroid.

**3 EXPERIMENTAL DEMONSTRATIONS**

The experimental system for SPGD algorithm demonstration is shown in Fig.1. The laser light (λ=632.8nm) is focused by a microscope objective and then pass a pinhole to simulate a point source. The collimated beam then passes through a transparent phase plate which induces static wavefront errors. Then the beam hit on a deformable mirror (DM) which has 140 active actuators within a clear aperture of 4.4 mm. This DM is based on Micro-Electro-Mechanical System (MEMS) technology and has a good initial surface quality (RMS error < 20nm). The wavefront after the DM is split to two parts by a beam splitter (BS). One part is imaged on the camera and the image data is acquired by the SPGD controller to command the DM. The other part goes to a Hartmann-Shark wavefront sensor (H-S WFS) that measures the wavefront error before and after correction for comparison.

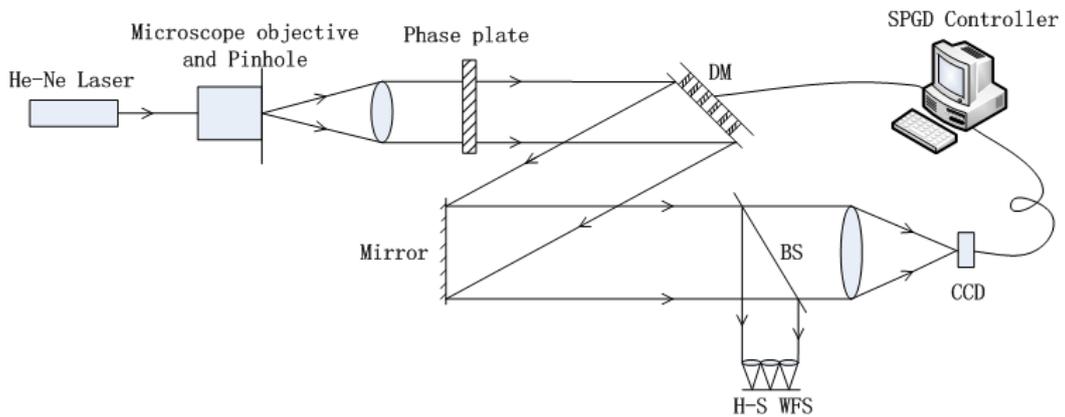

Fig.1 Experimental system for SPGD algorithm demonstration

The distorted wavefront caused by the phase plate and related far-field images before the DM correction are depicted in Fig.2, where the initial RMS wave-front error is 0.21λ. The wavefront and far-field image after correction are shown in Fig.3, where the RMS wave-front error is reduced to 0.041λ. The wavefront error is greatly reduced after the correction and the far-filed image is concentrated in a small circle. The evolution curve of metric in correction process is depicted in Fig.4. The metric achieve convergence in about 250 iterations in half minute. The elapsed time of the correction is dependent on the hardware configuration including actuator number, image acquisition speed and the software efficiency.

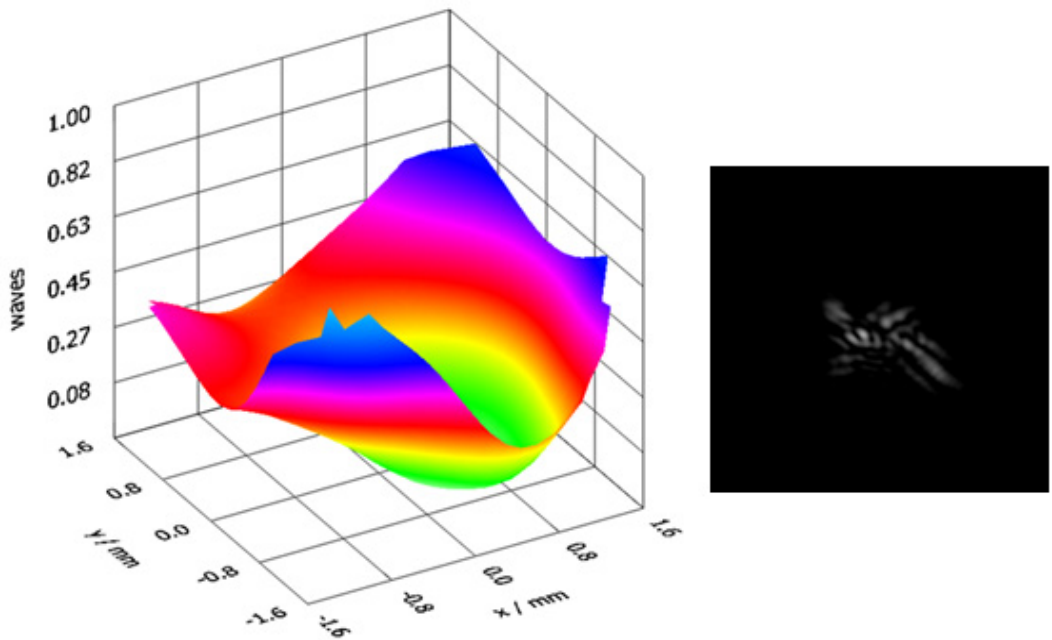

Fig.2 Distorted wavefront (PV=0.91λ; RMS=0.21λ) and far-field image (right)

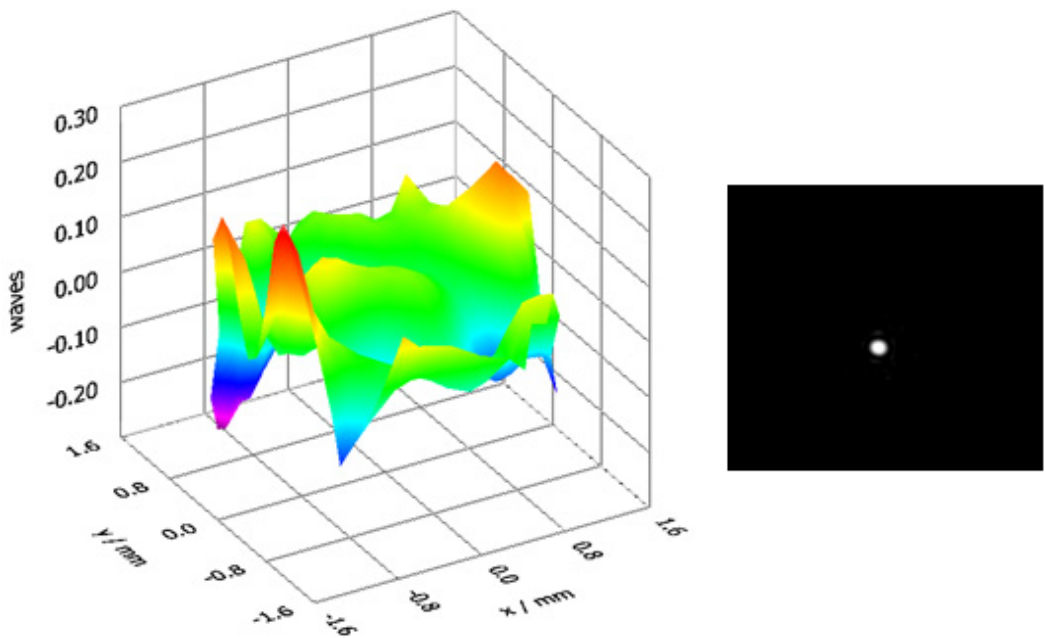

Fig.3 Corrected wavefront (PV=0.34λ; RMS=0.041λ) and far-field image (right)

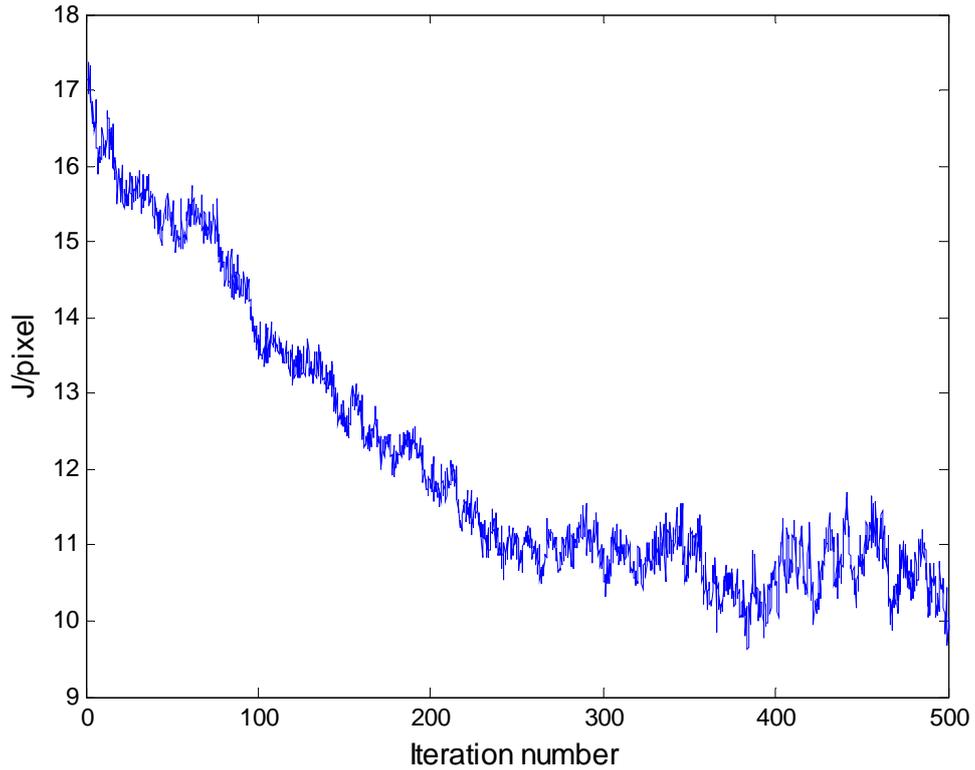

Fig.4 Evolution curve of metric J

It is easy to apply the SPGD method to the coronagraph based on liquid crystal array. The experimental lay-out is shown in Fig.5. The reflective liquid crystal array sandwiched by a polarizer and an analyzer serves as a spatial light modulator to modulate the amplitude of incoming light to a pre-designed pattern. Here a circular stepped pattern is generated by the liquid crystal array to form an apodized pupil (Ren & Zhu 2011, Dou et al. 2010). In this configuration, however, the H-S WFS cannot be used to obtain the phase map because of the pupil apodization.

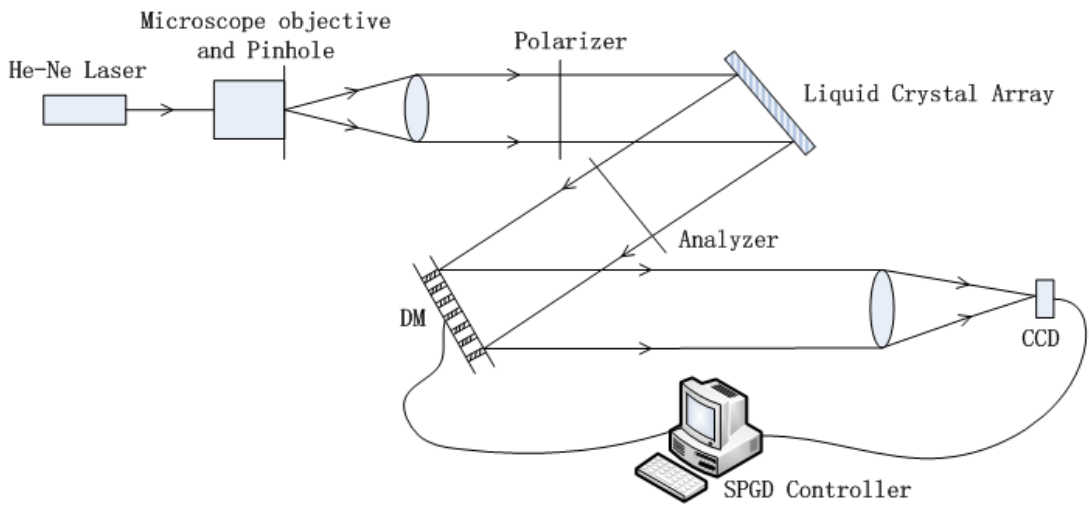

Fig.5 Layout of SPGD AO used in LCA based coronagraph

The apodized far-field images of coronagraph before and after the DM correction are shown in Fig.6. The image contrast is improved from $10^{-3}$ to $10^{-4.5}$ at an angular distance of larger than $2\lambda/D$ after correction, however, still not achieving its theoretical value of $10^{-5.5}$ (See Fig.7).

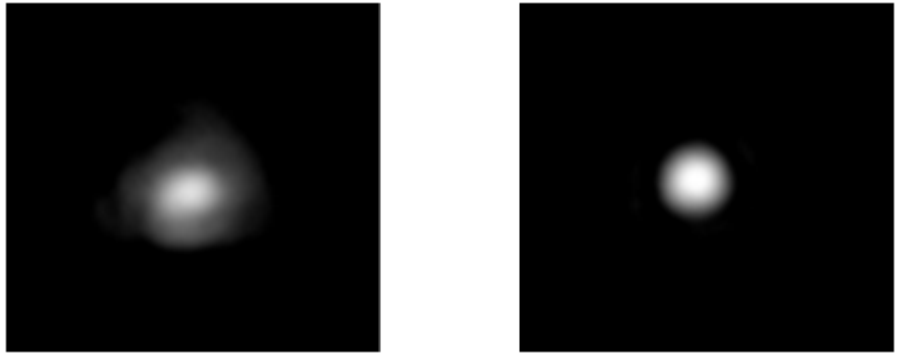

Fig.6 Apodized far-field image before (left) and after (right) correction

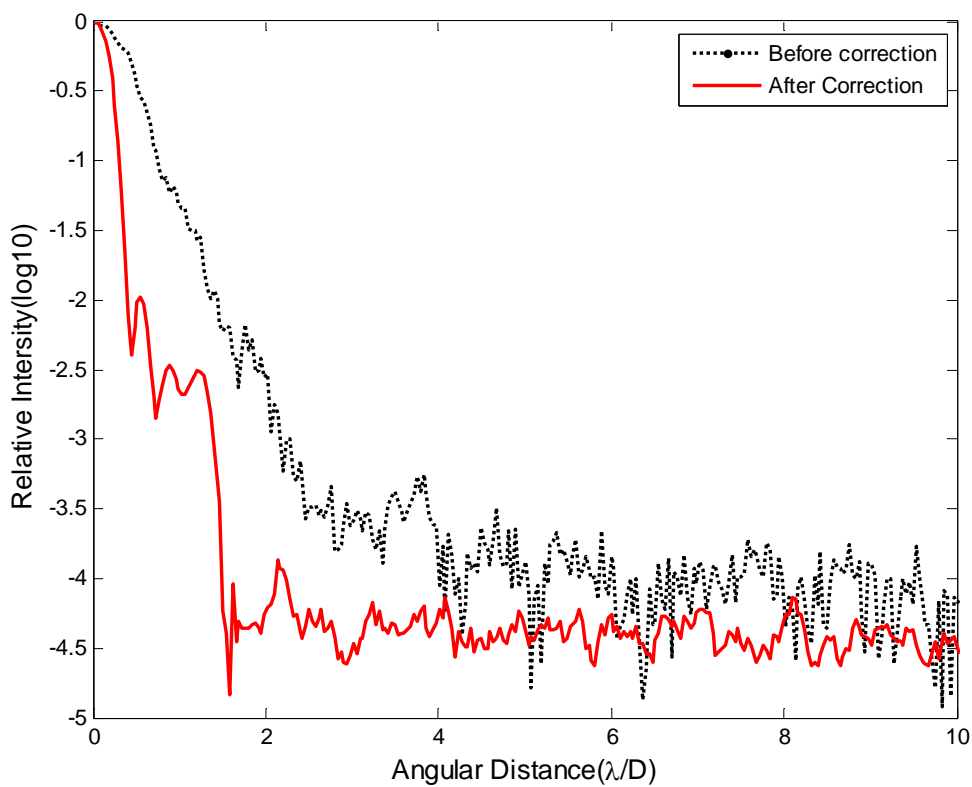

Fig.7 Contrast profile of coronagraph before and after AO correction

In high contrast imaging coronagraph application, we have to control the intensities in a wide dynamic range. The output of the camera currently used has only 8-bit dynamic range that is not enough to sample the low intensity areas some distance away from the PSF center, which leaves the intensities out of the main lobe undetected thus uncontrolled by SPGD optimization. Although we can extend the exposure time to acquire more image information, this will lead image saturation and unsteady of correction process. A possible solution is to use a camera with wider dynamic range like 16-bit or more. Also we can normalize several CCD images with different exposure time to one high-dynamic-range image which will be used to evaluate system metric in SPGD algorithm, however, sacrificing the speed.

Another factor that affects the performance of SPGD AO in speckle suppression of coronagraph is the actuator number of DM which determines the corrective ability in spatial frequency domain. In classical speckle nulling or "dark hole" algorithm, the actuator number of DM is usually larger than 1000 or more, which is an order of magnitude higher than ours. High actuator-density DM should be adopted in our system in the future.

We have made some efforts to calibrate the liquid crystal array to reduce its nonuniform response between different pixels, but it still has some impact on the image contrast to prevent it from approaching the design value.

## 4 CONCLUSIONS

SPGD based AO is a promising technique to eliminate the speckle noises and improve the contrast of imaging coronagraph. We have demonstrated the feasibility and good performance of SPGD based AO in laboratory and then applied it to the liquid crystal array based coronagraph, the image contrast is enhanced from $10^{-3}$ to $10^{-4.5}$ after correction. As the limitation of dynamic range of the detector, the actuator number of DM and the nonuniform response of the liquid crystal array, the image contrast has not come close to its design value and related problems should be discussed further in future publications.

## ACKNOWLEDGMENTS

This work was supported by the National Science Foundation under the grant ATM-0841440, the National Natural Science Foundation of China (NSFC) (Grant 10873024 and 11003031), as well as the National Astronomical Observatories' Special Fund for Astronomy-2009.